\documentstyle[prb,aps,floats,epsfig]{revtex}

\begin{document}
\draft
\twocolumn[\hsize\textwidth\columnwidth\hsize\csname @twocolumnfalse\endcsname

\title{Spectra and total energies from self-consistent many-body
perturbation theory}
\author{Arno Schindlmayr$^*$ and Thomas J. Pollehn$^\dagger$}
\address{Cavendish Laboratory, University of Cambridge, Madingley Road,
Cambridge CB3 0HE, United Kingdom}
\author{R. W. Godby}
\address{Department of Physics, University of York, Heslington, York YO10
5DD, United Kingdom}
\date{Received 9 June 1998}
\maketitle

\begin{abstract}
With the aim of identifying universal trends, we compare fully
self-consistent electronic spectra and total energies obtained from the
$GW$ approximation with those from an extended $GW\Gamma$ scheme that
includes a nontrivial vertex function and the fundamentally distinct
Bethe--Goldstone approach based on the $T$-matrix. The self-consistent
Green's function $G$, as derived from Dyson's equation, is used not only in
the self-energy but also to construct the screened interaction $W$ for a
model system. For all approximations we observe a similar deterioration of
the spectrum, which is not removed by vertex corrections. In particular,
satellite peaks are systematically broadened and move closer to the
chemical potential. The corresponding total energies are universally
raised, independent of the system parameters. Our results therefore suggest
that any improvement in total energy due to self-consistency, such as for
the electron gas in the $GW$ approximation, may be fortuitous.
\end{abstract}

]
\narrowtext

\section{Introduction}

Thanks to advances in modern computer technology and an increasingly
efficient treatment of the underlying one-electron structure, many-body
corrections to the quasiparticle band energies and spectral functions of
solids can now be obtained from first principles using many-body
perturbation theory. Most calculations for real materials employ the $GW$
approximation,\cite{Hed65} which owes its name to the fact that it models
the electron self-energy as the product $\Sigma^{GW} = i G W$ of the
Green's function $G$ and the dynamically screened Coulomb interaction $W$.
By explicitly including polarization effects in the exchange term it
describes dynamic correlation between the electrons and so can be
physically motivated as an extension of the static Hartree--Fock treatment.

The Green's function of the interacting electron system is linked to the
self-energy by means of Dyson's equation, symbolically written as $G^{-1} =
G^{{\rm H}^{\scriptstyle -1}} - \Sigma$, where $G^{\rm H}$ indicates the
Hartree approximation that neglects both exchange and correlation. It is
immediately clear that one faces a self-consistency problem, because the
self-energy in turn depends on the Green's function. Hence both propagators
must be determined simultaneously. The latter functional dependence is of
course nonlinear due to the dynamic properties of the screened interaction,
which is related to the bare Coulomb potential $v$ and the polarizability
$P$ through $W^{-1} = v^{-1} - P$. In a manner consistent with the $GW$
approximation the neglect of vertex corrections in the polarizability
yields the random-phase approximation $P^{\rm RPA} = -2 i G G$, which
ignores the interaction between the screening electrons and holes.

To obtain full self-consistency the above four equations have to be solved
iteratively starting from a zeroth-order noninteracting Green's function
until the results stabilize. Although self-consistent $GW$ calculations for
real materials are now within reach,\cite{Sch98a} the associated
computational cost is still enormous. Therefore in practice the outcome of
the first iteration is instead taken as the final spectrum. In this
formulation the $GW$ approximation has been applied to a wide range of
materials including semiconductors\cite{Hyb85,God86,Lin88} and alkali
metals\cite{Nor87} as well as transition metals\cite{Ary92} and their
oxides.\cite{Ary95} For all these diverse systems the predicted
quasiparticle band structures agree very well with experimental results,
while optical spectra, which include satellite features resulting from
collective excitations such as plasmons, are generally less satisfactory
and require the addition of so-called vertex corrections. However,
systematic progress in this direction is still limited.\cite{Sch98}

Despite the apparent success of conventional calculations, the neglect of
self-consistency remains problematic, in part because it implies a certain
ambiguity with respect to the choice of starting point. The zeroth-order
Green's function is usually constructed from the local-density
approximation (LDA), but in principle it is equally possible to start from
any other initial approximation such as the Hartree--Fock
treatment.\cite{Lin88} The resulting spectra will in general
differ.\cite{Pol98} Furthermore, the non-self-consistent $GW$ scheme
violates the Baym--Kadanoff criteria for conserving
approximations.\cite{Bay61} As a result the total particle number, energy,
and momentum of the system are not conserved under the influence of
external perturbations. Even without such perturbations, the integrated
spectral weight no longer corresponds to the number of physical
particles.\cite{Sch97}

In order to address these issues, past implementations have occasionally
incorporated modifications aimed at introducing a higher degree of
self-consistency. In particular, the band energies of the zeroth-order
Green's function used to evaluate the self-energy are sometimes shifted
such as to improve agreement with those obtained from Dyson's
equation.\cite{Hyb85,Nor87,Ary95} This approach assumes that the true
quasiparticle orbitals are virtually indistinguishable from the
corresponding LDA wave functions, which has only been explicitly proven for
states close to the band edge of simple semiconductors,
however.\cite{Hyb85} Moreover, it entirely ignores the transfer of spectral
weight to satellite peaks, which typically account for between 10\% and
50\% of the total spectrum.

More properly self-consistent results for model systems were recently
reported, although most realizations still restrict the computational
expense by fixing the screening function $W$ either at the zeroth-order
random-phase approximation\cite{Roh97,Bar96,Hol97} or a simpler
plasmon-pole model.\cite{Shi96} Until now the only comprehensive, fully
self-consistent calculations have been performed for a
quasi-one-dimensional semiconducting wire\cite{Far97,Gro95} and the
homogeneous electron gas.\cite{Hol98,Egu98} For the electron gas, the
system most studied so far,\cite{Bar96,Hol97,Shi96,Hol98,Egu98}
self-consistency was found to {\em worsen\/} the agreement between
calculated spectra and exact results by (i) increasing the occupied
bandwidth, (ii) transfering weight from the plasmon satellites to the
corresponding quasiparticle peaks, (iii) narrowing the quasiparticle
resonance widths, thereby increasing the lifetime, and (iv) broadening the
plasmon satellites while moving them closer to the Fermi surface. Some of
these effects have also been observed for the quasi-one-dimensional
wire,\cite{Gro95} and there is evidence that the reported increase in the
band gap extends to real semiconductors.\cite{Sch98a} In contrast,
self-consistency {\em improves\/} the agreement of quasiparticle energies
of localized semicore states with experimental data.\cite{Roh97}

Because of the small number of models studied so far the results quoted
above cannot readily be assumed for other systems without further
quantitative investigations, nor is it clear whether they are peculiar to
the $GW$ approximation or of a more general nature. Previous partially
self-consistent calculations that include vertex corrections have done
little to clarify the situation, since they only consider modifications of
the $GW$ scheme in the form of additional self-energy diagrams of second
order in $W$: depending on the choice of diagrams and the model screening
function used, these may restore the occupied band width of the electron
gas to its superior non-self-consistent value\cite{Shi96} or leave it
unchanged.\cite{Hol97}

In order to shed more light on these numerical aspects, in this paper we
present fully self-consistent calculations for a model system using a wide
range of conserving self-energy approximations. Besides the $GW$
approximation and an extended $GW\Gamma$ scheme that is derived from
time-dependent Hartree--Fock rather than Hartree theory and includes
multiple particle--hole scattering,\cite{Sch70} we also consider the
fundamentally distinct Bethe--Goldstone approach\cite{Bet57} based on the
$T$-matrix. Our first objective is to compare the resulting spectra and
thereby identify universal trends.

In the second part of this paper we then focus on total energies. A very
interesting outcome of recent fully self-consistent calculations for the
electron gas was that the total energy derived from the Green's function is
strikingly close to values obtained from quantum Monte-Carlo
simulations,\cite{Hol98} which are presumed accurate. It has been
speculated that this unexpected result is related to the fact that the
self-consistent $GW$ scheme conserves energy,\cite{Ary98} but the basis of
this connection is not immediately obvious. Rather, we will show here that
self-consistency in fact systematically raises the total energy. Our
results therefore suggest that the improvement for the electron gas may be
fortuitous.

This paper is organized as follows. In Sec.\ \ref{sec:model} we present the
model system and its exact numerical solution. In Sec.\ \ref{sec:approx} we
discuss the self-energy approximations considered here in more detail. In
Secs.\ \ref{sec:spectra} and \ref{sec:totalener} we give results for
spectral functions and total energies, respectively. Finally, in Sec.\
\ref{sec:conclusions} we summarize our conclusions.

\section{Model description}\label{sec:model}

In order to limit the computational cost of fully self-consistent
calculations with vertex corrections beyond the $GW$ approximation, which
so far have never been attempted for real materials, we consider a Hubbard
model that describes the dynamics of electrons on a lattice with strong,
short-range interaction. The Hamiltonian is sufficiently simple that it can
be diagonalized exactly for small cluster sizes using standard numerical
techniques, yet its physical behavior is nontrivial and reflects many
properties of real materials. The model geometry we employ is a finite
two-leg ladder with open boundary conditions. Each of the $M$ lattice sites
contains one orbital that can accommodate up to two electrons with opposite
spin. Doubly occupied orbitals are penalized by a repulsive on-site
interaction $U$, while the hopping of transient electrons between
neighboring sites yields an energy gain of $-t$. The full Hamiltonian is
\begin{equation} \label{eq:Hubbard}
{\cal H} = - t \sum_{\langle{\bf R},{\bf R}'\rangle,\sigma} c^\dagger_{{\bf
R}\sigma} c_{{\bf R}'\sigma} + U \sum_{\bf R}\hat{n}_{{\bf R}\uparrow}
\hat{n}_{{\bf R}\downarrow} ,
\end{equation}
where $c^\dagger_{{\bf R}\sigma}$, $c_{{\bf R}\sigma}$ are the creation and
annihilation operators for an electron at site ${\bf R}$ with spin
$\sigma$, $\hat{n}_{{\bf R}\sigma} \equiv c^\dagger_{{\bf R}\sigma}
c_{{\bf R}\sigma}$ is the particle number operator, and $\langle {\bf R},
{\bf R}' \rangle$ indicates a sum over nearest neighbors only. We choose
the energy norm by setting $t = 1$. The total electron number is denoted by
$N$.

The exact one-particle Green's function at zero temperature is defined as
\begin{equation} \label{eq:gexact}
G_{{\bf RR}'}(t-t') = -i \langle N | {\cal T} \{ c_{{\bf R}\sigma}(t)
c^\dagger_{{\bf R}'\sigma}(t') \} | N \rangle ,
\end{equation}
where $| N \rangle$ is the ground state of the interacting many-electron
system, ${\cal T}$ is Wick's time-ordering operator, and
$c_{{\bf R}\sigma}(t) \equiv \exp(i{\cal H}t) c_{{\bf R}\sigma}
\exp(-i{\cal H}t)$ denotes the time-dependent wave-field operator in the
Heisenberg picture. We have suppressed the spin index in $G$ because the
Green's function is diagonal and degenerate in $\sigma$. 

The Green's function can in principle be written in terms of the
eigenstates featuring an additional electron or hole,\cite{Fet71} but this
representation is disadvantageous because the basis set grows exponentially
with the system size. While the Hamiltonian matrix contains mostly zeroes
and so may be stored in a compressed format, the same is not possible for
the eigenvector matrix since it is not in general sparse. For a reasonable
system size the memory requirements thus render this procedure infeasible.
Instead, we Fourier transform (\ref{eq:gexact}) to the energy domain and
rewrite the Green's function in the form
\begin{eqnarray}
G_{{\bf RR}'}(\omega)
&=& \langle N | c_{{\bf R}\sigma} \frac{1}{\omega - {\cal H}^+ + E_N +
i\delta} c^\dagger_{{\bf R}'\sigma} | N \rangle \nonumber \\
&&+ \langle N | c^\dagger_{{\bf R}'\sigma} \frac{1}{\omega + {\cal H}^- -
E_N - i\delta} c_{{\bf R}\sigma} | N \rangle .
\end{eqnarray}
Here $E_N$ is the ground-state energy corresponding to $| N \rangle$, which
we compute by simultaneous subspace iteration,\cite{Par80} and
${\cal H}^\pm$ denotes the Hamiltonian matrix for $N \pm 1$ electrons. The
parameter $\delta$ is positive and tends to zero. In practice we use a
finite but small value of $\delta = 0.05$. The diagonal elements of $G$, in
which we are most interested, may now be calculated without full matrix
inversion by transforming $\omega \mp ({\cal H}^{\pm} - E_N - i\delta)$ to
a chain using the recursion method\cite{Hay80} and starting with the
vector $c^\dagger_{{\bf R}'\sigma} | N \rangle$ or $c_{{\bf R}\sigma} | N
\rangle$. Once the diagonal elements $a_n$ and off-diagonal elements $b_n$
of the tridiagonal matrix are determined up to a suitable chain length $D$,
the elements of the Green's function are obtained from
\begin{equation}
G_{\bf RR}(\omega) = \frac{1}{\displaystyle \omega - a_0 -
\frac{b_1^2}{\displaystyle \omega - a_1 - \cdots -
\frac{b_D^2}{\displaystyle \omega - a_D}}} .
\end{equation}
For nondiagonal elements of $G$ an analogous block recursion must be
performed. As the Hamiltonians considered here have many highly degenerate
eigenvalues, the chain length $D$ can be chosen substantially lower than
the order of ${\cal H}^\pm$. In practice a few recursions per actual
spectral feature are sufficient to achieve full convergence. To check the
accuracy we have also calculated the total particle number for all systems
discussed in the following by summing the diagonal elements of $G$ and
integrating the spectral weight below the chemical potential $\mu$. The
numerical deviation from the exact values is of the order of 0.1\%.

\section{Self-energy approximations}\label{sec:approx}

In many-body perturbation theory the effect of the Coulomb force on the
propagation of quasiparticles is rigorously described by an effective
potential. Following established conventions we distinguish between the
Hartree contribution
\begin{equation}
V^{\rm H}_{{\bf RR}'} = U ( \langle \hat{n}_{{\bf R}\uparrow} \rangle +
\langle \hat{n}_{{\bf R}\downarrow} \rangle ) \delta_{{\bf RR}'}
\end{equation}
and the remaining exchange--correlation part, which we call the self-energy
$\Sigma$. It is in general both nonlocal and energy-dependent. Although the
exact self-energy functional remains elusive, physically motivated
approximations can be obtained by truncating its diagrammatic series
expansion. In the following we describe the three distinct schemes
considered in this paper.

\begin{figure}[b!]
\epsfxsize=3.125in \centerline{\epsfbox{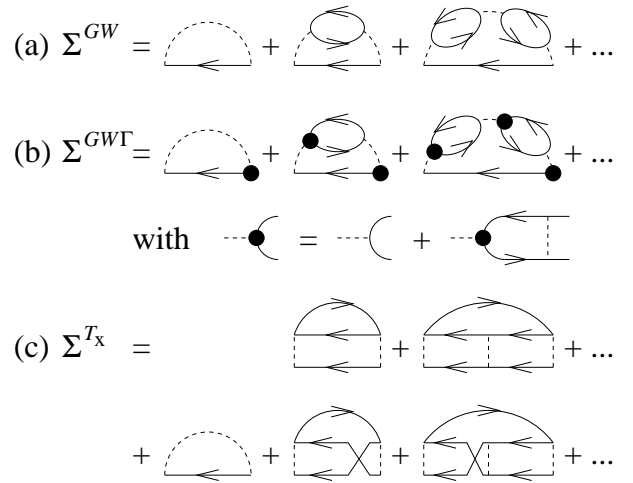}} \medskip
\caption{Diagrammatic representation of (a) the $GW$ approximation, (b) a
$GW\Gamma$ scheme with vertex corrections that describe multiple
particle--hole scattering, and (c) the Bethe--Goldstone approach based on
the $T$-matrix. Arrows represent Green's functions; the Coulomb interaction
is indicated by a broken line.}\label{fig:approx}
\end{figure}

The $GW$ approximation renormalizes the nonlocal Fock potential by
including dynamic screening in the exchange interaction, as shown
diagrammatically in Fig.\ \ref{fig:approx}(a). The screening function is
modeled in the random-phase approximation (RPA) and includes ring diagrams
to all orders. In the spirit of the space--time method\cite{Roj95} we avoid
costly convolutions by switching between the real time and energy domains
as appropriate, using fast Fourier transforms with 32,768 sampling points
over a range of 160 energy units. This procedure also guarantees a high
degree of numerical accuracy, because we do not need to repeatedly fit the
propagators to analytic functions. Given a Green's function $G$ we hence
compute the self-energy by solving the defining equations 
\begin{eqnarray}
P^{\rm RPA}_{{\bf RR}'}(t) &=& -2 i G_{{\bf RR}'}(t)
G_{{\bf R}'{\bf R}}(-t) , \\
W^{-1}_{{\bf RR}'}(\omega) &=& \frac{1}{U} \delta_{{\bf RR}'} -
P^{\rm RPA}_{{\bf RR}'}(\omega) , \\
\Sigma^{GW}_{{\bf RR}'}(t) &=& i G_{{\bf RR}'}(t) W_{{\bf RR}'}(t) .
\label{eq:sigmagw}
\end{eqnarray}
The factor 2 in the polarization propagator is due to spin summation.

While the $GW$ approximation accurately describes materials that are
governed by the screening of free carriers, such as the homogeneous
electron gas, vertex corrections are in general necessary for more complex
systems. Such extensions are often refered to as $GW\Gamma$ schemes. A
particular approximation that we consider here includes a vertex function
derived from time-dependent Hartree--Fock theory, as shown in Fig.\
\ref{fig:approx}(b). It contains multiple scattering in the particle--hole
channel, which is most significant in atomic and molecular systems with
partially filled shells.\cite{Shi93} Nontrivial vertex functions usually
increase the computational cost dramatically, but due to the short-range
interaction in the Hubbard model the self-energy in this case is still
given by an expression of the form (\ref{eq:sigmagw}), albeit with a
modified screened interaction
\begin{equation}
\tilde{W}^{-1}_{{\bf RR}'}(\omega) = \frac{1}{U} \delta_{{\bf RR}'} -
\frac{1}{2} P^{\rm RPA}_{{\bf RR}'}(\omega) .
\end{equation}

The Bethe--Goldstone approach constitutes a fundamentally distinct
approximation based on the so-called transition or $T$-matrix, which
describes multiple scattering in the particle--particle and hole--hole
channels to all orders. This process dominates in the low-density limit of
the electron gas,\cite{Kan63} but it also predicts the specific behavior of
systems with localized orbitals and strong electronic correlation such as
the transition metals.\cite{Spr98} As the self-energy, shown as a sum of
ladder diagrams in Fig.\ \ref{fig:approx}(c), contains exchange
contributions in the two-particle propagator, we designate it by the label
$T_{\rm x}$. For the Hubbard model the corresponding direct and exchange
terms are in fact identical except for a prefactor of 2 due to spin
summation in the former, so the self-energy is given by
\begin{eqnarray}
G^2_{{\bf RR}'}(t) &=& i G_{{\bf RR}'}(t) G_{{\bf RR}'}(t) , \\
T^{-1}_{{\bf RR}'}(\omega) &=& \frac{1}{U} \delta_{{\bf RR}'} -
G^2_{{\bf RR}'}(\omega) , \\
\Sigma^{T_{\rm x}}_{{\bf RR}'}(t) &=& -i T_{{\bf RR}'}(t)
G_{{\bf R}'{\bf R}}(-t) - V^{\rm H}_{{\bf RR}'} \delta(t) .
\end{eqnarray}
In the last equation we have subtracted the Hartree potential because it is
already dealt with separately.

By tuning the parameters of the Hamiltonian (\ref{eq:Hubbard}) we can
create configurations geared to the particular strengths of different
self-energy approximations within the same model: independent of the
Coulomb integral $U$ the $T$-matrix becomes increasingly accurate for a
very low or, because of particle--hole symmetry, very high fractional band
filling $N/(2M)$, while the $GW$ schemes perform best for medium site
occupancies and a not too strong interaction.\cite{Pol98}

\section{Self-consistent spectra}\label{sec:spectra}

In order to study the effects of self-consistency in a general perspective,
we compare calculations using all the many-body approximations described in
the previous section. As a convenient starting point we choose the Hartree
Green's function $G^{\rm H}$, which only includes the electrostatic
potential $V^{\rm H}$ generated by the total electron charge. The
occupation numbers $\langle \hat{n}_{{\bf R}\sigma} \rangle$ are determined
self-consistently by a simple iterative procedure. After evaluating the
self-energy $\Sigma$ we obtain an updated, dressed Green's function from
Dyson's equation
\begin{equation} \label{eq:Dyson}
G^{-1}_{{\bf RR}'}(\omega) = G^{{\rm H}^{\scriptstyle -1}}_{{\bf RR}'}
(\omega) - \Sigma_{{\bf RR}'}(\omega) ,
\end{equation}
which in a conventional treatment is taken as the final spectrum. In a
self-consistent calculation we instead use it to compute a new Hartree
potential and self-energy and continue the iteration until the results
stabilize. To guarantee the correct analytic time-ordering of the spectrum
obtained from Dyson's equation it is necessary to shift the Hartree Green's
function rigidly on the energy axis by an amount $\langle
\Sigma(\mu^{\rm H}) \rangle$ before evaluating the self-energy in
(\ref{eq:Dyson}). Here $\mu^{\rm H}$ denotes the chemical potential, which
we identify with the highest occupied quasiparticle state, and the matrix
element is formed with the corresponding orbital. As this shift must tend
to zero for the self-consistent solution, we decrease it by a factor of
$e^{-1}$ in every subsequent iteration. To achieve convergence we typically
perform at least ten iterations, after which the shift is reduced to a
negligible value without influence on the spectral features or total
energy. We use a very small initial resonance width of $\delta = 0.05$
throughout the calculations in order to avoid systematic errors, and only
the final spectra are broadened through convolution with a Lorentzian of
width 0.5 for visual display. We have again checked the numerical
reliability by calculating the total particle number from the
self-consistent Green's functions and generally find the same high level of
accuracy as for the exact solution.

\begin{figure}[b!]
\epsfxsize=3.125in \centerline{\epsfbox{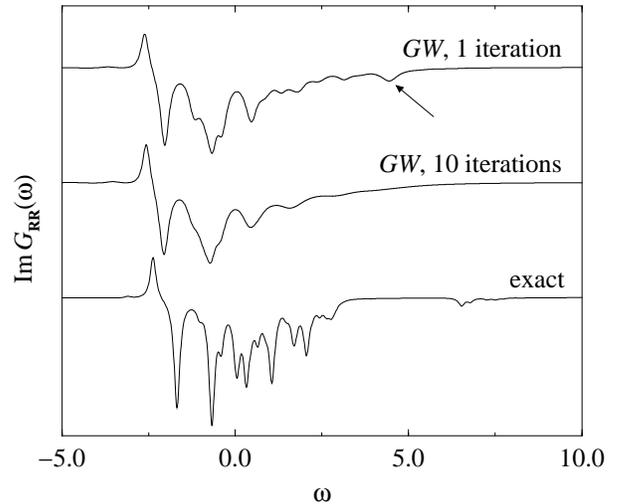}}
\caption{Comparison between an ordinary and a converged $GW$ calculation
after 10 iterations. The most striking effect of self-consistency is the
broadening of satellite peaks, which are hardly discernible in a diffuse
background. This is particularly obvious at high energies, as indicated by
an arrow.}\label{fig:selfconsGW}
\end{figure}

In Fig.\ \ref{fig:selfconsGW} we compare an ordinary $GW$ spectral
function, obtained from a single iteration of Dyson's equation, with the
result of a converged, self-consistent calculation after 10 iterations.
Like all other figures in this section it shows the diagonal element for a
corner site of the cluster, which we have confirmed to be representative.
The following observations therefore apply equally to other matrix
elements. By setting the model parameters to $M = 10$ and $N = 2$ with a
medium interaction strength of $U = 4$ we have deliberately chosen a small
band filling of 10\% for which the $GW$ approximation is not optimal, so
that possible improvements should be more obvious. The exact spectrum is
shown for comparison.

\begin{figure}
\epsfxsize=3.125in \centerline{\epsfbox{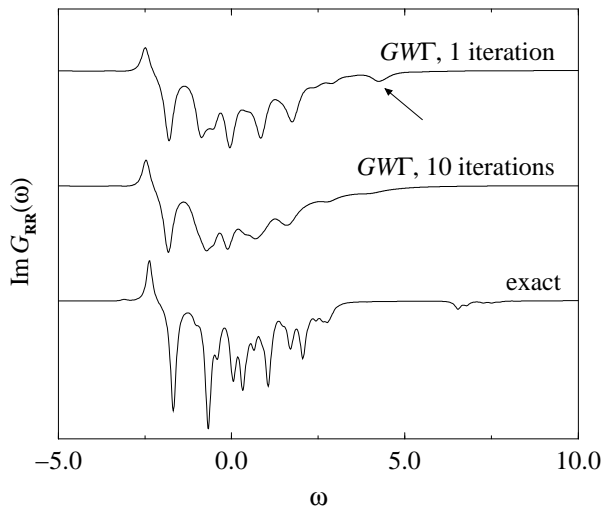}}
\caption{The vertex function in the $GW\Gamma$ approximation fails to
prevent the deterioration of spectral features when brought to full
self-consistency.}\label{fig:selfconsGWGamma}
\end{figure}

\begin{figure}[b!]
\epsfxsize=3.125in \centerline{\epsfbox{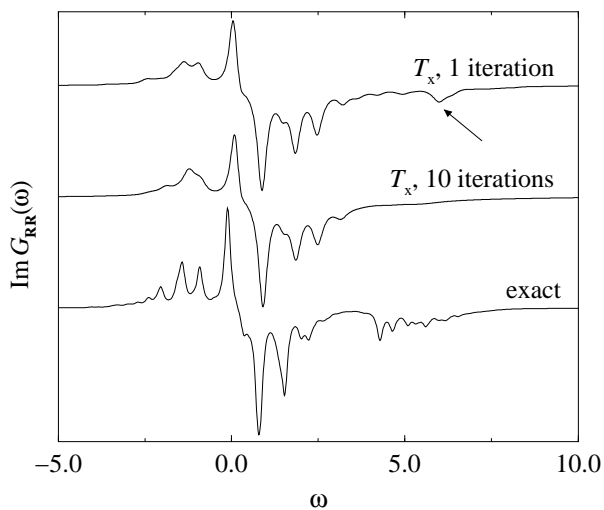}}
\caption{Although unrelated to the $GW$ approximation, the $T_{\rm x}$
scheme effects the same changes in quasiparticle and satellite peaks in a
converged, self-consistent spectrum.}\label{fig:selfconsTx}
\end{figure}

To examine the effects of self-consistency we distinguish between
quasiparticles and their satellites. The former are in fact little
affected, mainly by small shifts in position and a marginal narrowing of
the resonance width for states close to the chemical potential. In
contrast, the satellite spectrum deteriorates significantly. The most
striking change is the broadening of satellite peaks. By fitting the
spectrum to a set of Lorentzians we find that the resonance widths
approximately double. As the spectral weight is smeared out and peaks
merge, individual features are hardly discernible above the spectral
background, particularly isolated satellites at high energies. One such
example is indicated by an arrow. Furthermore, it can be seen that the
satellites move closer towards the chemical potential. All of these
observations are in agreement with previous self-consistent
calculations.\cite{Bar96,Hol98}

We now address the relation between self-consistency and the simultaneous
inclusion of vertex corrections. To this end we show the results of a
corresponding $GW\Gamma$ calculation in Fig.\ \ref{fig:selfconsGWGamma}. It
is evident that the vertex function in the latter fails to prevent the
deterioration of spectral features when brought to full self-consistency.
In particular, we do not observe a restoration of well-defined structure.
This point is further underlined by our investigation of the
Bethe--Goldstone approach. A typical calculation using $N = 6$, which
corresponds to a band filling of 30\%, is illustrated in Fig.\
\ref{fig:selfconsTx}: although unrelated to the $GW$ approximation, the
$T_{\rm x}$ scheme effects the same changes in quasiparticle and satellite
peaks in a converged, self-consistent spectrum.

The general success of ordinary $GW$ calculations for most systems suggests
that the sum of all neglected self-energy diagrams is small. A thorough
understanding of this process is invaluable for the design of superior
approximations, but despite continuing efforts the nature of this
cancellation remains elusive. On the one hand, there is strong evidence for
a certain mutual cancellation between corresponding vertex corrections in
the polarizability and self-energy.\cite{Mah89,Ver95,Bec97} If we assume
that this argument still holds for the true vertex function, then by
extension the remaining contributions, i.e., the self-consistent
renormalization of propagators in the polarizability and self-energy, must
also cancel. However, the disturbing deterioration of spectral features in
the self-consistent $GW$ approximation,\cite{Hol98} particularly when
compared to a partially self-consistent calculation with a fixed
zeroth-order dielectric function,\cite{Bar96} suggests that this is not the
case, at least when a trivial vertex is used. Consequently one
would also expect a certain mutual cancellation between self-consistency
and vertex diagrams. Recently reported direct numerical
evidence\cite{Shi96} to this end is circumstantial, however, since only
quasiparticle properties were considered and the response function was
replaced by a plasmon-pole model of unclear diagrammatic structure. In this
context our results, alongside those from a partially self-consistent
cumulant expansion that also examined complete spectral
functions,\cite{Hol97} indicate that this cancellation is a very subtle
process and that mutually balancing contributions may be hard to identify.

\section{Total energies}\label{sec:totalener}

One of the notable features of self-consistency in many-body perturbation
theory is that the total energy becomes a proper, uniquely defined
quantity. In practice total energies are most often obtained from the
one-particle Green's function using the Galitskii--Migdal
formula,\cite{Gal58} which may of course be applied to any approximate $G$.
However, it is important to note that other definitions, for instance
through the two-particle Green's function or an integral of the interaction
strength, in general yield different numerical values. Full
self-consistency removes this ambiguity. Moreover, the total energy is then
also properly conserved under the influence of external
perturbations.\cite{Bay61}

For the Hubbard model an analogous expression for the total energy
\begin{equation} \label{eq:totalener}
E = \frac{1}{\pi} \sum_{\langle{\bf R},{\bf R}'\rangle}
\int_{-\infty}^\mu\!\! \left( \omega \delta_{{\bf RR}'} - t \right)
\mbox{Im}\, G_{{\bf RR}'}(\omega) \,d{\omega}
\end{equation}
in terms of the one-particle Green's function can be derived from the
equation of motion of the wave-field operator. Despite formal similarities
to the Galitskii--Migdal formula this expression also contains
contributions from nondiagonal elements of the Green's function. The reason
for this apparent discrepancy is that the site index ${\bf R}$ does not
represent a spatial coordinate. Instead, it is introduced in second
quantization to label a set of overlapping Wannier orbitals. As the
creation and annihilation operators are transformed separately, local
operators in real space become nondiagonal in the site index, a point that
was previously noted in conjunction with the proper parametrization of the
charge density.\cite{Sch95} Analogously the local kinetic-energy operator
and external potential correspond to the nondiagonal hopping term of the
Hubbard Hamiltonian, which resurfaces here.

Unlike the calculation of the particle number, a direct evaluation of the
frequency integral in (\ref{eq:totalener}) proved quite sensitive to the
initial resonance width $\delta$. In order to obtain reliable results, we
therefore fit the elements of the Green's function to a model of the form
\begin{eqnarray} \label{eq:fit}
G_{{\bf RR}'}(\omega)
&=& \sum_n \frac{a_{{\bf RR}'}^n}{\omega - b_{{\bf RR}'}^n - i
\delta_{{\bf RR}'}^n} \nonumber \\
&&+ \sum_m \frac{a_{{\bf RR}'}^m}{\omega - b_{{\bf RR}'}^m + i
\delta_{{\bf RR}'}^m} ,
\end{eqnarray}
which of course becomes exact as $\delta$ tends to zero. The frequency
integration can now be performed analytically in the proper limit
$\delta_{{\bf RR}'}^n \to 0$, and so for the total energy we eventually
obtain
\begin{equation}
E = \sum_{\langle{\bf R},{\bf R}'\rangle} \sum_n \left( b_{{\bf RR}'}^n
\delta_{{\bf RR}'} - t \right) a_{{\bf RR}'}^n .
\end{equation}
We have confirmed the reliability of our procedure by comparing total
energies derived from the true Green's function with the exact numerical
value $E_N$, which we obtained earlier by diagonalizing the Hamiltonian
matrix. The fit according to (\ref{eq:fit}) is very accurate. Unfortunately
it is also computationally expensive, so that all results in this section
refer to a reduced model size of $M = 6$.

\begin{figure}
\epsfxsize=3.125in \centerline{\epsfbox{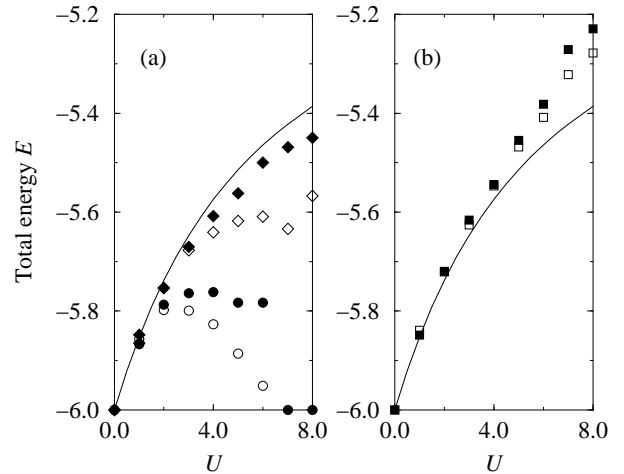}}
\caption{Total energies calculated from (a) the $GW$ and $GW\Gamma$
approximations, indicated by circles and diamonds, respectively, and (b)
the $T_{\rm x}$ scheme as a function of the interaction strength $U$ for a
low band filling of 17\%. Open symbols refer to ordinary,
non-self-consistent Green's functions, while filled symbols refer to
self-consistent ones. The solid line shows the exact total energy for
comparison.}\label{fig:totalener_N2}
\end{figure}

\begin{figure}[b!]
\epsfxsize=3.125in \centerline{\epsfbox{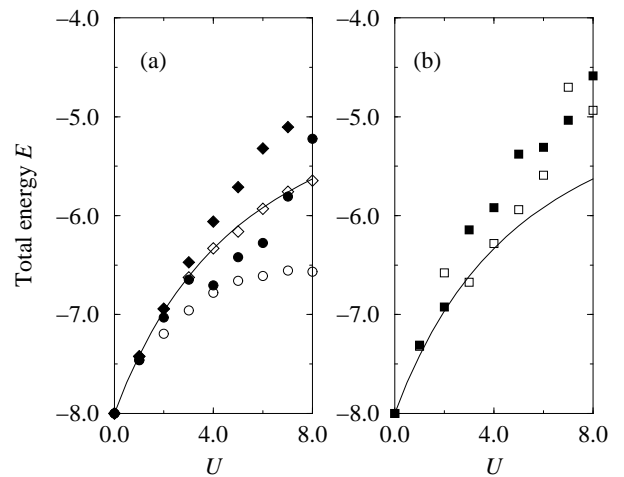}}
\caption{Corresponding values for the same model with an intermediate band
filling of 33\%. Self-consistency systematically raises the total energy
for all approximations.}\label{fig:totalener_N4}
\end{figure}

In Figs.\ \ref{fig:totalener_N2} and \ref{fig:totalener_N4} we show
calculated total energies as a function of the interaction strength $U$ for
$N = 2$, which corresponds to a low band filling of 17\%, and $N = 4$,
equivalent to an intermediate band filling of 33\%. Results obtained
from the $GW$, the $GW\Gamma$, and the $T_{\rm x}$ scheme are indicated by
circles, diamonds, and squares, respectively. Open symbols refer to
ordinary, non-self-consistent Green's functions, while filled symbols refer
to self-consistent ones. The curves are not perfectly smooth due to
unresolved convergence problems for individual values of $U$. As these do
not obscure overall trends, however, we have decided to retain the
corresponding energies for reasons of completeness. The solid line shows
the exact total energy for comparison.

As a first result we note that the quality of total-energy predictions
correlates with that of spectral functions in the same parameter range,
i.e., the Bethe--Goldstone approach works best for low particle numbers,
while the two $GW$ schemes perform optimally for intermediate band
fillings, where screening effects dominate. Of the latter, the $GW\Gamma$
approximation is superior due to the prominence of exchange in the Hubbard
model. The drop in the $GW$ total energy after reaching a plateau is in
part an artefact of the model specification: as the interaction is
short-range, the true energy converges to a finite value in the limit $U
\to \infty$ as long as $N \leq M$, indicating a complete spatial separation
of the electrons. The tendency of the $GW$ approximation to underestimate
the total energy eventually causes the downward trend. Such an unphysical
behavior does not occur for the long-range Coulomb interaction, where the
energy diverges as the correlation strength approaches infinity.

\begin{figure}[b!]
\epsfxsize=3.125in \centerline{\epsfbox{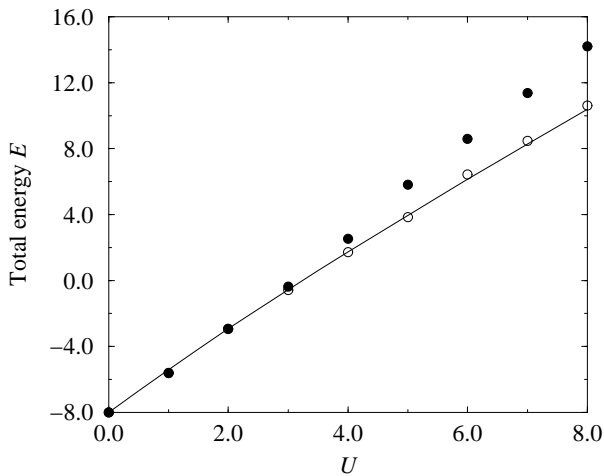}}
\caption{Total energy in the $GW$ approximation for a larger band filling
of 67\%. As the ordinary $GW$ approximation already predicts the true
energies rather well in this case, self-consistency leads to a substantial
overestimation.}\label{fig:totalener_N8}
\end{figure}

When comparing the total energies obtained from ordinary with those
obtained from self-consistent Green's functions, we find that the latter
are systematically raised. This is a general feature valid for all
approximations at all band fillings that we investigated. It can be
understood as follows: self-consistency modifies the Green's function in
two ways, namely by rescaling and moving individual resonances relative to
the chemical potential, and by an overall rigid shift caused by a
redefinition of the chemical potential itself. The first effect may
influence the total energy in either way, depending on the balance of
opposite trends. For the homogeneous electron gas, for instance, the
increase in band width, which moves quasiparticles to lower energies
relative to the Fermi surface, competes with the upward transfer of
spectral weight from low-lying plasmon satellites to the corresponding main
peaks.\cite{Hol98} In contrast, the second contribution is always positive.
It results from a relocation of the chemical potential, which in an
ordinary treatment is given by $\mu = \mu^{\rm H} + \langle
\Sigma(\mu^{\rm H}) \rangle$, equivalent to that of the shifted Hartree
Green's function in Dyson's equation. In a self-consistent approach the
chemical potential instead becomes $\mu = \mu^{\rm H} + Z \langle
\Sigma(\mu^{\rm H}) \rangle$, where $Z$ is the quasiparticle
renormalization factor. For the sake of the argument we ignore small
deviations in the underlying Hartree potentials and the self-energy matrix
elements, which are of course calculated from different Green's functions.
It is then clear that in the self-consistent case the self-energy
correction, which is always negative, is scaled down by $Z$, leading to a
higher reference chemical potential. Unless compensated by other factors,
this effect therefore always raises the total energy.

Although the argument makes the increase in total energy for the electron
gas plausible, it does not explain the excellent numerical agreement with
results from quantum Monte-Carlo simulations, which are presumed accurate.
In the light of our calculations this appears rather fortuitous, however.
While the increase in the $GW$ total energy for our model system with two
and four electrons indeed constitutes a quantitative improvement, in Fig.\
\ref{fig:totalener_N8} we present results for a larger band filling of
67\%, corresponding to $N = 8$, to demonstrate that this is not always so:
as the ordinary $GW$ approximation already predicts the true energies
rather well in this case, self-consistency leads to a substantial
overestimation.

\section{Conclusions}\label{sec:conclusions}

In this paper we have presented self-consistent many-body calculations of
the spectra and total energies for a model system. The self-consistency was
not restricted and extends to the construction of the screened interaction
in the random-phase approximation. By comparing the $GW$ approximation with
an extended $GW\Gamma$ scheme that includes a nontrivial vertex function
as well as the unrelated Bethe--Goldstone approach based on the $T$-matrix
we were able to identify universal trends. We have demonstrated that the
deterioration of spectral features due to self-consistency, previously
observed in $GW$ calculations, also occurs in more elaborate treatments and
is not removed by vertex corrections. The most important effect is the
broadening of satellite peaks, particularly at high energies, and their
simultaneous shift towards the chemical potential. For all approximations
the corresponding total energies are systematically raised. This trend,
which we made plausible on the basis of physical arguments, is independent
of the system parameters such as correlation strength and band filling. Our
results therefore suggest that the recently reported improvement in the
$GW$ total energy for the electron gas due to self-consistency may be
fortuitous.

\acknowledgements

We are grateful to R.\ Haydock and C.~M.~M.\ Nex for helpful discussions.
This work was supported by the Royal Society and the European Community
program Human Capital and Mobility through contract No.\ CHRX-CT93-0337.
A.\ Schindl\-mayr and T.~J.\ Pollehn wish to thank the
Stu\-di\-en\-stif\-tung des deut\-schen Vol\-kes for financial support. A.\
Schindl\-mayr gratefully acknowledges further support from the Deut\-scher
Aka\-de\-mi\-scher Aus\-tausch\-dienst under its HSP III scheme, the
Gott\-lieb Daim\-ler- und Karl Benz-Stif\-tung, Pembroke College Cambridge,
and the Engineering and Physical Sciences Research Council.


\begin{references}
\bibitem[*]{schindlmayr} Electronic address: as10031@phy.cam.ac.uk
\bibitem[\dagger]{pollehn} Present address: The Boston Consulting Group,
Sendlinger Stra{\ss}e 7, 80331 M\"unchen, Germany.
\bibitem{Hed65} L. Hedin, Phys. Rev. {\bf 139}, A796 (1965).
\bibitem{Sch98a} W.-D. Sch\"one and A. G. Eguiluz (unpublished).
\bibitem{Hyb85} M. S. Hybertsen and S. G. Louie, Phys. Rev. Lett. {\bf 55},
1418 (1985); Phys. Rev. B {\bf 32}, 7005 (1985); {\bf 34}, 5390 (1986).
\bibitem{God86} R. W. Godby, M. Schl\"uter, and L. J. Sham, Phys. Rev.
Lett. {\bf 56}, 2415 (1986); Phys. Rev. B {\bf 35}, 4170 (1986); {\bf 37},
10~159 (1988).
\bibitem{Lin88} W. von der Linden and P. Horsch, Phys. Rev. B {\bf 37},
8351 (1988).
\bibitem{Nor87} J. E. Northrup, M. S. Hybertsen, and S. G. Louie, Phys.
Rev. Lett. {\bf 59}, 819 (1987); Phys. Rev. B {\bf 39}, 8198 (1989); M. P.
Surh, J. E. Northrup, and S. G. Louie, {\it ibid.} {\bf 38}, 5976 (1988).
\bibitem{Ary92} F. Aryasetiawan and U. von Barth, Physica Scripta {\bf
T45}, 270 (1992); F. Aryasetiawan, Phys. Rev. B {\bf 46}, 13~051 (1992).
\bibitem{Ary95} F. Aryasetiawan and O. Gunnarsson, Phys. Rev. Lett.
{\bf 74}, 3221 (1995).
\bibitem{Sch98} A. Schindlmayr and R. W. Godby, Phys. Rev. Lett. {\bf 80},
1702 (1998).
\bibitem{Pol98} T. J. Pollehn, A. Schindlmayr, and R. W. Godby, J. Phys.:
Condens. Matter {\bf 10}, 1273 (1998).
\bibitem{Bay61} G. Baym and L. P. Kadanoff, Phys. Rev. {\bf 124}, 287
(1961).
\bibitem{Sch97} A. Schindlmayr, Phys. Rev. B {\bf 56}, 3528 (1997).
\bibitem{Roh97} M. Rohlfing, P. Kr\"uger, and J. Pollmann, Phys. Rev. B
{\bf 56}, R7065 (1997).
\bibitem{Bar96} U. von Barth and B. Holm, Phys. Rev. B {\bf 54}, 8411
(1996); {\bf 55}, 10~120(E) (1997).
\bibitem{Hol97} B. Holm and F. Aryasetiawan, Phys. Rev. B {\bf 56}, 12~825
(1997).
\bibitem{Shi96} E. L. Shirley, Phys. Rev. B {\bf 54}, 7758 (1996).
\bibitem{Far97} B. Farid, Philos. Mag. B {\bf 76}, 145 (1997).
\bibitem{Gro95} H. J. de Groot, P. A. Bobbert, and W. van Haeringen, Phys.
Rev. B {\bf 52}, 11~000 (1995).
\bibitem{Hol98} B. Holm and U. von Barth, Phys. Rev. B {\bf 57}, 2108
(1998).
\bibitem{Egu98} A. G. Eguiluz and W.-D. Sch\"one, Mol. Phys. {\bf 94}, 87
(1998).
\bibitem{Sch70} B. Schneider, H. S. Taylor, and R. Yaris, Phys. Rev. A
{\bf 1}, 855 (1970).
\bibitem{Bet57} H. A. Bethe and J. Goldstone, Proc. R. Soc. London
{\bf A238}, 551 (1957).
\bibitem{Ary98} F. Aryasetiawan and O. Gunnarsson, Rep. Prog. Phys.
{\bf 61}, 237 (1998).
\bibitem{Fet71} A. L. Fetter and J. D. Walecka, {\it Quantum Theory of
Many-Particle Systems} (McGraw--Hill, San Francisco, 1971).
\bibitem{Par80} B. N. Parlett, {\it The Symmetric Eigenvalue Problem}
(Prentice Hall, Englewood Cliffs, 1980).
\bibitem{Hay80} R. Haydock, in {\it Solid State Physics}, edited by H.
Ehrenreich, F. Seitz, and D. Turnbull (Academic, New York, 1980), Vol. 35,
p. 215.
\bibitem{Roj95} H. N. Rojas, R. W. Godby, and R. J. Needs, Phys. Rev. Lett.
{\bf 74}, 1827 (1995).
\bibitem{Shi93} E. L. Shirley and R. M. Martin, Phys. Rev. B {\bf 47},
15~404 (1993).
\bibitem{Kan63} J. Kanamori, Prog. Theor. Phys. {\bf 30}, 275 (1963).
\bibitem{Spr98} M. Springer and F. Aryasetiawan, Phys. Rev. Lett. {\bf 80},
2389 (1998).
\bibitem{Mah89} G. D. Mahan and B. E. Sernelius, Phys. Rev. Lett. {\bf 62},
2718 (1989).
\bibitem{Ver95} C. Verdozzi, R. W. Godby, and S. Holloway, Phys. Rev. Lett.
{\bf 74}, 2327 (1995).
\bibitem{Bec97} F. Bechstedt, K. Tenelsen, B. Adolph, and R. Del Sole,
Phys. Rev. Lett. {\bf 78}, 1528 (1997).
\bibitem{Gal58} V. M. Galitskii and A. B. Migdal, Sov. Phys. JETP {\bf 7},
96 (1958).
\bibitem{Sch95} A. Schindlmayr and R. W. Godby, Phys. Rev. B {\bf 51},
10~427 (1995).
\end{references}
\end{document}